\documentclass[aps,prb,twocolumn,showpacs]{revtex4}

\usepackage{graphicx}
\usepackage{amsmath}
\usepackage{amssymb}

\begin{document}

\title{Momentum dependence of quasiparticle spectrum and Bogoliubov
angle in cuprate superconductors}

\author{Weifang Wang, Zhi Wang, Jingge Zhang, and Shiping Feng$^{*}$}

\affiliation{Department of Physics, Beijing Normal University,
Beijing 100875, China}

\begin{abstract}
The momentum dependence of the low energy quasiparticle spectrum and
the related Bogoliubov angle in cuprate superconductors are studied
within the kinetic energy driven superconducting mechanism. By
calculation of the ratio of the low energy quasiparticle spectra at
positive and negative energies, it is shown that the Bogoliubov
angle increase monotonically across the Fermi crossing point. The
results also show that the superconducting coherence of the low
energy quasiparticle peak is well described by a simple d-wave
Bardeen-Cooper-Schrieffer formalism, although the pairing mechanism
is driven by the kinetic energy by exchanging spin excitations.
\end{abstract}


\maketitle

After over twenty years of extensive studies, an agreement has
emerged that superconductivity in cuprate superconductors results
when electrons pair up into Cooper pairs \cite{tsuei} as in the
conventional superconductors \cite{bcs}, then these electron Cooper
pairs condensation reveals the superconducting (SC) ground-state.
However, as a natural consequence of the unconventional SC mechanism
that is responsible for the high SC transition temperatures
\cite{anderson}, the electron Cooper pairs in cuprate
superconductors have a dominant d-wave symmetry \cite{tsuei,shen}.
However, in spite of the unconventional SC mechanism, the
angle-resolved photoemission spectroscopy (ARPES) experimental
results have unambiguously established the Bogoliubov quasiparticle
nature of the sharp SC quasiparticle peak in cuprate superconductors
\cite{matsui,campuzano1}, then the SC coherence of the low energy
quasiparticle peak is well described by a simple d-wave
Bardeen-Cooper-Schrieffer (BCS) formalism \cite{bcs}. In the
framework of the BCS formalism, the Bogoliubov quasiparticle is a
coherent combination of particle (electron) and its absence (hole),
i.e., its annihilation operator is a linear combination of particle
and hole operators as \cite{bogoliubov,schrieffer} $\gamma_{{\bf
k}\uparrow}=U_{\bf k} c_{{\bf k}\uparrow}+V_{\bf k} c^{\dag}_{-{\bf
k} \downarrow}$, with the constraint for the coherence factors
$|U_{\bf k}|^2+|V_{\bf k} |^2=1$ for any wave vector $\bf k$
(normalization). In this case, the Bogoliubov quasiparticle do not
carry definite charge. This particle-hole dualism of Bogoliubov
quasiparticles then is responsible for a variety of profound
phenomena in the SC state. In particular, the coherence factors
$U_{\bf k}$ and $V_{\bf k}$ for cuprate superconductors as a
function of momentum have been determined experimentally from the
ARPES measurements \cite{matsui}.

Recently a quantity referred to as the Bogoliubov angle has been
introduced in terms of the coherence factors $U_{\bf k}$ and $V_{\bf
k}$ as \cite{balatsky,fujita},
\begin{eqnarray}\label{angle}
\Theta_{{\bf k}}=\arctan\left (\left [{|U_{\bf k}|^2\over |V_{\bf
k}|^2} \right]^{1/2} \right ),
\end{eqnarray}
which is the manifestation of the particle-hole dualism of the SC
quasiparticles, for example, for $\Theta_{{\bf k}}=0$ the Bogoliubov
quasiparticle excitation will be a hole-like, whereas in the
opposite case of $\Theta_{{\bf k}}=\pi/2$ the Bogoliubov
quasiparticle is essentially an electron-like. Moreover, the angle
that corresponds to the strongest admixture between particle and
hole is $\Theta_{{\bf k}}=\pi/4$. In a simple BCS formalism
\cite{schrieffer}, this Bogoliubov angle can be rewritten as,
\begin{eqnarray}\label{angle1}
\Theta_{{\bf k}}=\arctan\left [{A({\bf k},\omega>0)\over A({\bf k},
\omega <0)}\right ]^{1/2},
\end{eqnarray}
obviously, it is a new observable quantity closely related to the
ARPES spectrum of superconductors \cite{balatsky,fujita}. Moreover,
this Bogoliubov angle reflects the relative weight of particle and
hole amplitudes in the Bogoliubov quasiparticle, and therefore plays
an essential role in characterizing the SC state via quantities such
as the SC gap and its symmetry. By comparing the ratio of the ARPES
spectral intensities at positive and negative energies, the momentum
dependence of the Bogoliubov angle for cuprate superconductors has
been extensively studied \cite{balatsky}. In particular, these ARPES
experimental studies \cite{matsui,balatsky} for the momentum
dependence of the coherence factors and the related Bogoliubov angle
have further confirmed the validity of the basic d-wave BCS
formalism description of the SC state in cuprate superconductors,
and therefore have also introduced important constraints on the
microscopic model and SC mechanism. To the best of our knowledge,
the momentum dependence of the coherence factors and the related
Bogoliubov angle has not been treated starting from a microscopic SC
theory, and no explicit calculations of the evolution of the
particle and hole mixing has been made so far.

In this paper, we study the momentum dependence of the low energy
quasiparticle spectrum and the related Bogoliubov angle for cuprate
superconductors based on the kinetic energy driven SC mechanism
\cite{feng1}. We employed the $t$-$J$ model, and then show
explicitly that the Bogoliubov angle for cuprate superconductors
increase monotonically across the Fermi crossing point.

We start from the two-dimensional $t$-$J$ model on a square lattice
\cite{anderson,shen},
\begin{eqnarray}\label{t-j1}
H&=&-t\sum_{i\hat{\eta}\sigma}C^{\dagger}_{i\sigma}
C_{i+\hat{\eta}\sigma}+t'\sum_{i\hat{\tau}\sigma}
C^{\dagger}_{i\sigma}C_{i+\hat{\tau}\sigma}+\mu\sum_{i\sigma}
C^{\dagger}_{i\sigma}C_{i\sigma}\nonumber\\
&+&J\sum_{i\hat{\eta}}{\bf S}_{i} \cdot {\bf S}_{i+\hat{\eta}},
\end{eqnarray}
acting on the Hilbert subspace with no double occupied site, i.e.,
$\sum_{\sigma}C^{\dagger}_{i\sigma}C_{i\sigma}\leq 1$, where
$\hat{\eta}=\pm\hat{x},\pm \hat{y}$, $\hat{\tau}=\pm\hat{x}
\pm\hat{y}$, $C^{\dagger}_{i\sigma}$ ($C_{i\sigma}$) is the electron
creation (annihilation) operator, ${\bf S}_{i}=(S^{x}_{i},S^{y}_{i},
S^{z}_{i})$ are spin operators, and $\mu$ is the chemical potential.
To deal with the constraint of no double occupancy in analytical
calculations, the charge-spin separation (CSS) fermion-spin theory
\cite{feng2,feng3} has been developed, where the constrained
electron operators are decoupled as $C_{i\uparrow}=
h^{\dagger}_{i\uparrow}S^{-}_{i}$ and $C_{i\downarrow}=
h^{\dagger}_{i\downarrow}S^{+}_{i}$, with the spinful fermion
operator $h_{i\sigma}=e^{-i\Phi_{i\sigma}}h_{i}$ describes the
charge degree of freedom together with some effects of the spin
configuration rearrangements due to the presence of the doped hole
itself (charge carrier), while the spin operator $S_{i}$ describes
the spin degree of freedom (spin), then the electron local
constraint for single occupancy is satisfied in analytical
calculations. In particular, it has been shown that under the
decoupling scheme, this CSS fermion-spin representation is a natural
representation of the constrained electron defined in the Hilbert
subspace without double electron occupancy \cite{feng3}. In the CSS
fermion-spin representation, the $t$-$J$ Hamiltonian (\ref{t-j1})
can be expressed as,
\begin{eqnarray}\label{t-j2}
H&=&t\sum_{i\hat{\eta}}(h^{\dagger}_{i+\hat{\eta}\uparrow}
h_{i\uparrow}S^{+}_{i}S^{-}_{i+\hat{\eta}}+
h^{\dagger}_{i+\hat{\eta} \downarrow}h_{i\downarrow}S^{-}_{i}
S^{+}_{i+\hat{\eta}})\nonumber\\
&-&t'\sum_{i\hat{\tau}} (h^{\dagger}_{i+\hat{\tau}\uparrow}
h_{i\uparrow}S^{+}_{i} S^{-}_{i+\hat{\tau}}+
h^{\dagger}_{i+\hat{\tau}\downarrow}h_{i\downarrow}S^{-}_{i}
S^{+}_{i+\hat{\tau}})\nonumber \\
&-&\mu\sum_{i\sigma}h^{\dagger}_{i\sigma} h_{i\sigma}+J_{{\rm eff}}
\sum_{i\hat{\eta}}{\bf S}_{i}\cdot {\bf S}_{i+\hat{\eta}},
\end{eqnarray}
with $J_{{\rm eff}}=(1-\delta)^{2}J$, and $\delta=\langle
h^{\dagger}_{i\sigma}h_{i\sigma}\rangle=\langle h^{\dagger}_{i}
h_{i}\rangle$ is the hole doping concentration. As a consequence,
the kinetic energy term in the $t$-$J$ model has been transferred as
the interaction between charge carriers and spins, which reflects
that even the kinetic energy term in the $t$-$J$ Hamiltonian has a
strong Coulombic contribution due to the restriction of no double
occupancy of a given site.

For the understanding of the physical properties of cuprate
superconductors in the SC state, we have developed a kinetic energy
driven SC mechanism \cite{feng1}, where the interaction between
charge carriers and spins from the kinetic energy term in the
$t$-$J$ model (\ref{t-j2}) induces the charge carrier pairing state
with the d-wave symmetry by exchanging spin excitations, then the
electron Cooper pairs originating from the charge carrier pairing
state are due to the charge-spin recombination, and their
condensation reveals the SC ground-state. In particular, this d-wave
SC state is controlled by both the SC gap function and the
quasiparticle coherence, which leads to a fact that the maximal SC
transition temperature occurs around the optimal doping, and then
decreases in both underdoped and overdoped regimes \cite{feng1}.
Furthermore, it has been shown that this SC state is a conventional
BCS-like with the d-wave symmetry \cite{guo,feng3}, so that the
basic BCS formalism with the d-wave SC gap function is still valid
in discussions of the low energy electronic structure of cuprate
superconductors, although the pairing mechanism is driven by the
kinetic energy by exchanging spin excitations, and other exotic
magnetic scattering \cite{dai} is beyond the d-wave BCS formalism.
Following the previous discussions \cite{guo,feng3,feng1}, the
charge carrier diagonal and off-diagonal Green's functions can be
obtained as,
\begin{eqnarray}
g({\bf k},\omega)&=&Z_{hF}\left ({U^{2}_{h{\bf k}}\over \omega-
E_{h{\bf k}}}+{V^{2}_{h{\bf k}} \over \omega+E_{h{\bf k}}}\right ),
\label{gg1}\\
\Im^{\dagger}({\bf k},\omega)&=&-Z_{hF}{\bar{\Delta}_{hZ}({\bf k})
\over 2E_{h{\bf k}}}\left ({1\over \omega-E_{h{\bf k}}}-{1\over
\omega+ E_{h{\bf k}}}\right ),~~~\label{gg2}
\end{eqnarray}
where the charge carrier quasiparticle spectrum $E_{h{\bf k}}=\sqrt
{\bar{\xi^{2}_{{\bf k}}}+\mid \bar{\Delta}_{hZ}({\bf k})\mid^{2}}$
with the renormalized d-wave charge carrier pair gap function
$\bar{\Delta}_{hZ}({\bf k})=\bar{\Delta}_{hZ}[{\rm cos}k_{x}-{\rm
cos}k_{y}]/2$, and the charge carrier quasiparticle coherence
factors $U^{2}_{h{\bf k}} =(1+\bar{\xi_{{\bf k}}}/E_{h{\bf k}})/2$
and $V^{2}_{h{\bf k}}= (1-\bar{\xi_{{\bf k}}}/E_{h{\bf k}})/2$,
while the charge carrier quasiparticle coherent weight $Z_{hF}$ and
other notations are defined as same as in Ref. \onlinecite{guo}, and
have been determined by the self-consistent calculation
\cite{guo,feng3,feng1}.

In the CSS fermion-spin theory \cite{feng2}, the electron diagonal
and off-diagonal Green's functions are the convolutions of the spin
Green's function and charge carrier diagonal and off-diagonal
Green's functions in Eqs. (\ref{gg1}) and (\ref{gg2}), respectively.
Following the previous discussions \cite{guo}, we can obtain the
electron diagonal and off-diagonal Green's functions in the present
case, and then the electron spectral function from electron diagonal
Green's function is obtained as,
\begin{eqnarray}\label{AA}
A({\bf k},\omega)&=&2\pi {1\over N}\sum_{{\bf p}}Z_{F}{B_{{\bf p}}
\over 2\omega_{{\bf p}}}\nonumber\\
&\times&[U^{2}_{h{\bf p+k}}L_{1}({\bf k},{\bf p})
\delta(\omega+E_{h{\bf p+k}}-\omega_{{\bf p}})\nonumber\\
&+&U^{2}_{h{\bf p+k}} L_{2}({\bf k},{\bf p})\delta(\omega+E_{h{\bf
p+k}}+\omega_{{\bf p}})
\nonumber\\
&+&V^{2}_{h{\bf p+k}}L_{1}({\bf k},{\bf p}) \delta(\omega-E_{h{\bf
p+k}}+\omega_{{\bf p}})\nonumber\\
&+&V^{2}_{h{\bf p+k}}L_{2}({\bf k},{\bf p})\delta(\omega-E_{h{\bf
p+k}}- \omega_{{\bf p}})],~~~~~
\end{eqnarray}
where the electron quasiparticle coherent weight $Z_{F}=Z_{hF}/2$,
$L_{1}({\bf k},{\bf p})={\rm coth}[(\beta\omega_{{\bf p}})/2]-{\rm
tanh}[(\beta E_{h{\bf p+k}})/2]$ and $L_{2}({\bf k},{\bf p})={\rm
coth}[(\beta\omega_{{\bf p}})/2]+{\rm tanh}[(\beta E_{h{\bf p+k}})/2
]$, and the spin excitation spectrum $\omega_{{\bf p}}$ and $B_{\bf
p}$ have been given in Ref. \onlinecite{guo}. For the convenience of
discussions, the electron spectral function in Eq. (\ref{AA}) also
can be rewritten formally as,
\begin{eqnarray}\label{AA1}
A({\bf k},\omega)=2\pi Z_{F}[U^{2}_{{\bf k}}\delta(\omega-E_{{\bf k}
})+V^{2}_{{\bf k}}\delta(\omega+E_{{\bf k}})],
\end{eqnarray}
then the electron coherence factors $U^{2}_{{\bf k}}$ and
$V^{2}_{{\bf k}}$ can be obtained as,
\begin{eqnarray}
U^{2}_{{\bf k}}\delta(\omega-E_{{\bf k}})&=&{1\over N}\sum_{{\bf p}
}{B_{{\bf p}}\over 2\omega_{{\bf p}}}[L_{1}({\bf k},{\bf p})
\delta(\omega-E_{h{\bf p+k}}+\omega_{{\bf p}})\nonumber\\
&+&L_{2}({\bf k},{\bf p} )\delta(\omega-E_{h{\bf p+k}}
-\omega_{{\bf p}})]V^{2}_{h{\bf p+k}},\label{uu}\\
V^{2}_{{\bf k}}\delta(\omega+E_{{\bf k}})&=&{1\over N}\sum_{{\bf p}
}{B_{{\bf p}}\over 2\omega_{{\bf p}}}[L_{1}({\bf k},{\bf p})
\delta(\omega+E_{h{\bf p+k}}-\omega_{{\bf p}})\nonumber\\
&+&L_{2}({\bf k},{\bf p} )\delta(\omega+E_{h{\bf p+k}}+\omega_{{\bf
p}}]U^{2}_{h{\bf p+k}}, \label{vv}
\end{eqnarray}
with the electron quasiparticle spectrum $E_{{\bf k}}$. With the
help of the spectral function (\ref{AA}), the Bogoliubov angle in
the present case for cuprate superconductors is expressed explicitly
as,
\begin{eqnarray}\label{angle3}
\Theta_{{\bf k}}=\arctan\left [{A({\bf k},\omega>0)\over A({\bf k},
\omega <0)}\right ]^{1/2}.
\end{eqnarray}
In particular, this Bogoliubov angle (\ref{angle3}) can be used to
determined the Fermi surface as it has been done in the experiments
\cite{balatsky}. This follows from a fact that at the Fermi crossing
point $k_F$, the electron coherence factors
$U^{2}_{k_{F}}=V^{2}_{k_{F}}$, and then the Bogoliubov angle
$\Theta_{k_{F}}=\pi/4$.

\begin{figure}[t]
\begin{center}
\leavevmode
\includegraphics[clip=true,width=0.8\columnwidth]{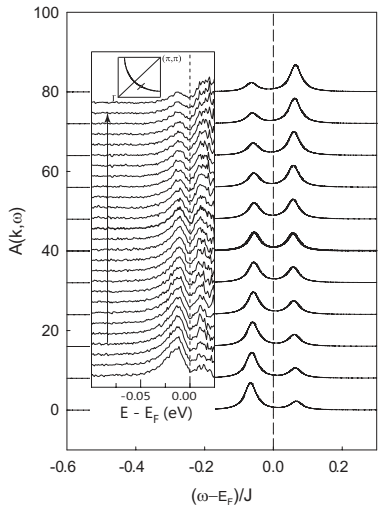}
\caption{The quasiparticle spectral function along the cut position
$[0.82\pi,0.57\pi]$ to $[0.83\pi,0.58\pi]$ with $T=0.002J$ for
$t/J=2.5$ and $t'/t=0.3$ at $\delta=0.15$. Inset: the corresponding
experimental results taken from Ref. \onlinecite{balatsky}.}
\end{center}
\end{figure}

\begin{figure}[t]
\begin{center}
\leavevmode
\includegraphics[clip=true,width=0.6\columnwidth]{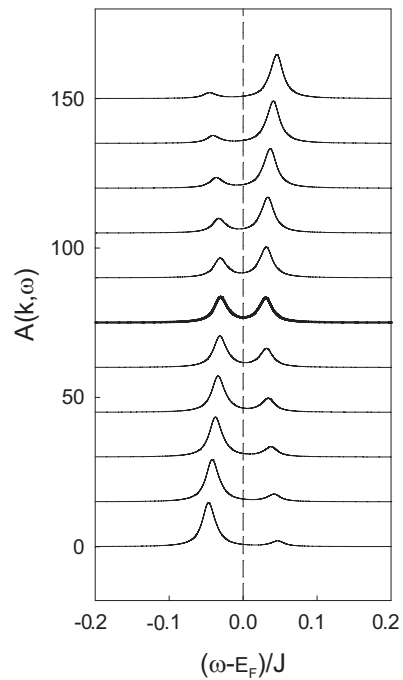}
\caption{The quasiparticle spectral function along the cut position
$[0.776\pi,0.651\pi]$ to $[0.786\pi,0.661\pi]$ with $T=0.002J$ for
$t/J=2.5$ and $t'/t=0.3$ at $\delta=0.12$.}
\end{center}
\end{figure}

In cuprate superconductors, although the values of $J$ and $t$ is
believed to vary somewhat from compound to compound
\cite{wells,tanaka}, however, as a qualitative discussion, the
commonly used parameters in this paper are chosen as $t/J=2.5$,
$t'/t=0.3$. We are now ready to discuss the energy and momentum
dependence of the SC quasiparticle spectral function $A({\bf k},
\omega)$ in Eq. (\ref{AA}) and the related Bogoliubov angle
$\Theta_{{\bf k}}$ in Eq. (\ref{angle3}). In Fig. 1, we plot $A({\bf
k}, \omega)$ as a function of energy along the cut direction
$[0.82\pi,0.57\pi]$ to $[0.83\pi,0.58\pi]$ crossing the Fermi
surface with temperature $T=0.002J$ at the doping concentration
$\delta=0.15$ in comparison with the corresponding experimental
result \cite{balatsky} for the optimally doped cuprate
superconductor Bi$_{2}$Sr$_{2}$CaCu$_{2}$O$_{8+\delta}$ (inset). The
thick solid curve is the momentum distribution curve where the
electron coherence factors $U^{2}_{{\bf k}}=V^{2}_{{\bf k}}$ and has
been defined as the Fermi crossing point just as it has been done in
the experiments \cite{balatsky}. Obviously, the spectral weight of
the two branches is momentum dependent. These two sharp
quasiparticle peaks in each energy distribution curve exhibit a
evolution of the relative peak height at different momentum
positions. Moreover, the quasiparticle spectral intensity of the two
bands show an opposite evolution as a function of ${\bf k}$ along
the cut position $[0.82\pi,0.57\pi]$ to $[0.83\pi,0.58\pi]$. This is
a common feature of the momentum dependence of the SC quasiparticle
spectral function along the cut direction crossing the Fermi
surface. For a better understanding of the momentum dependence of
the SC quasiparticle spectral function, we have made a series of
calculations for the momentum dependence of the SC quasiparticle
spectral function along different cut directions crossing the Fermi
surface at different doping concentration levels, and the result of
$A({\bf k},\omega)$ as a function of energy along the cut direction
$[0.776\pi,0.651\pi]$ to $[0.786\pi,0.661\pi]$ crossing the Fermi
surface with temperature $T=0.002J$ at the doping concentration
$\delta=0.12$ is plotted in Fig. 2. The quasiparticle peak below the
Fermi surface has a higher intensity than that above the Fermi
surface. However, after passing the Fermi surface, the quasiparticle
peak above the Fermi surface has a higher intensity than that below
the Fermi surface. This crossover behavior near the Fermi surface, a
characteristic of the Bogoliubov quasiparticle dispersion in the
conventional superconductors in the SC state, appears in cuprate
superconductors. To show this point clearly, we plot the
quasiparticle peak intensity along the cut direction
$[0.65\pi,0.4\pi]$ to $[\pi,0.75\pi]$ crossing the Fermi surface
with temperature $T=0.002J$ at the doping concentration
$\delta=0.15$ in Fig. 3. For comparison, the corresponding
experimental result \cite{balatsky} for the optimally doped cuprate
superconductor Bi$_{2}$Sr$_{2}$CaCu$_{2}$O$_{8+\delta}$ is also
plotted in Fig. 3 (inset). In comparison with the results in Fig. 1
and Fig. 2, we therefore confirm that (i) although the dispersive
feature in Fig. 1 and Fig.2 is almost symmetrical with respect to
the Fermi surface, the SC quasiparticle peak intensity is not; (ii)
the quasiparticle spectral intensity break near the Fermi surface
shows the existence of a gap and two branches of dispersion centered
at the Fermi surface; (iii) both bands show the bending back effect
at the Fermi surface. All these theoretical results are
qualitatively consistent with the ARPES experimental data of cuprate
superconductors \cite{matsui,balatsky}. Incorporating our previous
results \cite{guo}, we therefore confirming that the basic d-wave
BCS formalism under the kinetic energy driven SC mechanism can
correctly reproduce some low energy features of the SC coherence of
the quasiparticle peaks observed in cuprate superconductors
\cite{shen,matsui,balatsky}, including the doping and temperature
dependence of the electron spectral function at the antinodal point
and the momentum dependence of the electron spectral function along
the cut direction crossing the Fermi surface.

\begin{figure}[t]
\begin{center}
\leavevmode
\includegraphics[clip=true,width=0.8\columnwidth]{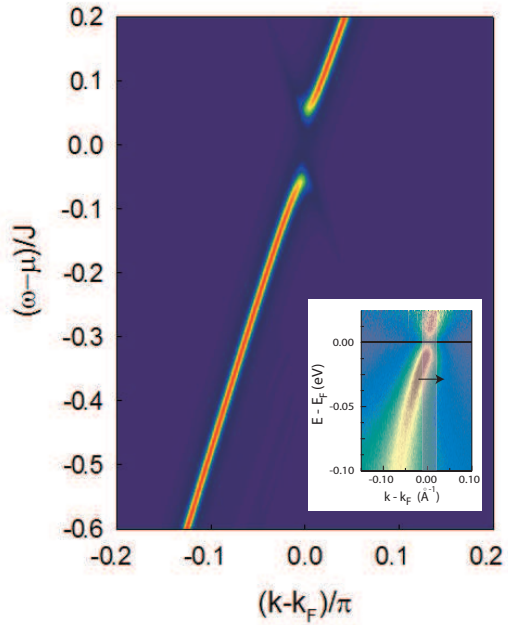}
\caption{A color plot of the quasiparticle peak intensity along the
cut position $[0.65\pi,0.4\pi]$ to $[\pi,0.75\pi]$ with $T=0.002J$
for $t/J=2.5$ and $t'/t=0.3$ at $\delta=0.15$. Inset: the
corresponding experimental results taken from Ref.
\onlinecite{balatsky}.}
\end{center}
\end{figure}

\begin{figure}[t]
\begin{center}
\leavevmode
\includegraphics[clip=true,width=0.8\columnwidth]{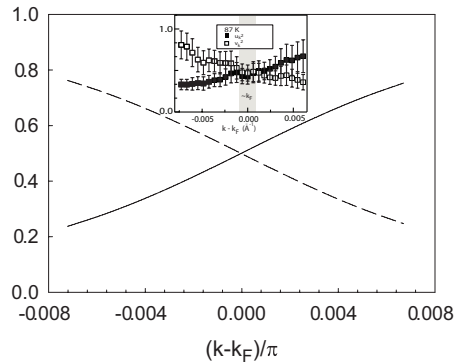}
\caption{The quasiparticle coherence factors $U^{2}_{{\bf k}}$
(solid line) and $V^{2}_{{\bf k}}$ (dashed line) along the cut
position $[0.82\pi,0.57\pi]$ to $[0.83\pi,0.58\pi]$ with $T=0.002J$
for $t/J=2.5$ and $t'/t=0.3$ at $\delta=0.15$. Inset: the
corresponding experimental results taken from Ref.
\onlinecite{balatsky}.}
\end{center}
\end{figure}

Now we turn to discuss the momentum dependence of the electron
coherence factors in Eq. (\ref{uu}) and Eq. (\ref{vv}) and the
related Bogoliubov angle Eq. (\ref{angle3}). From Eq. (\ref{AA1}),
we can find that the SC quasiparticle peak hight of the peak below
the Fermi surface in Fig. 1 and Fig. 2 is assigned a weight
$Z_{F}V^{2}_{{\bf k}}$, while that of the peak above the Fermi
surface is assigned a weight $Z_{F}U^{2}_{{\bf k}}$, therefore the
coherence factors describe the relative intensity of the Bogoliubov
quasiparticle bands above and below the Fermi surface. In Fig. 4, we
plot $U^{2}_{{\bf k}}$ (solid line) and $V^{2}_{{\bf k}}$ (dashed
line) along the cut direction $[0.82\pi,0.57\pi]$ to
$[0.83\pi,0.58\pi]$ with temperature $T=0.002J$ at the doping
concentration $\delta=0.15$ in comparison with the corresponding
experimental result \cite{balatsky} for the optimally doped cuprate
superconductor Bi$_{2}$Sr$_{2}$CaCu$_{2}$O$_{8+\delta}$ (inset),
where the particle-hole mixing takes place due to the pairing,
leading to a transfer of weight between the electron and hole bands.
In particular, the electron coherence factors $U^{2}_{{\bf k}}$ and
$V^{2}_{{\bf k}}$ have contrary evolution, and they are equivalent
at the Fermi wave vector $k_F$, then $V^{2}_{{\bf k}}+U^{2}_{{\bf
k}}=1$ is always satisfied, showing good agreement in the band
dispersion between the experiment \cite{matsui,balatsky} and the
present theoretical calculation. For a further confirmation of the
conventional Bogoliubov quasiparticle behaviors in cuprate
superconductors, we have employed the ratio of the low energy
quasiparticle spectra at positive and negative energies as a measure
of the Bogoliubov angle $\Theta_{{\bf k}}$ (\ref{angle3}) at each
momentum just as it has been done in the experiments
\cite{balatsky}. The result for the extracted the Bogoliubov angle
$\Theta_{{\bf k}}$ along the cut direction $[0.82\pi,0.57\pi]$ to
$[0.83\pi,0.58\pi]$ with temperature $T=0.002J$ at the doping
concentration $\delta=0.15$ is plotted in Fig. 5 in comparison with
the corresponding experimental result \cite{balatsky} for the
optimally doped cuprate superconductor
Bi$_{2}$Sr$_{2}$CaCu$_{2}$O$_{8+\delta}$ (inset). We therefore find
that $\Theta_{{\bf k}}$ increase monotonically across the Fermi
crossing point $k_F$ suggesting a continuously evolution of the
particle and hole mixing within this momentum range. As we expect,
$\Theta_{{\bf k}}=\pi/4$ at $k_F$, indicating that the particle and
hole mix equally at $k_F$ for cuprate superconductors, since in this
case the weight of the SC quasiparticle peak below the Fermi surface
$Z_{F}V^{2}_{{\bf k}}$ is the same as the weight of the peak above
the Fermi surface $Z_{F}U^{2}_{{\bf k}}$ as mentioned above. Our
this result is qualitatively consistent with the experimental data
for cuprate superconductors \cite{balatsky}.

\begin{figure}[t]
\begin{center}
\leavevmode
\includegraphics[clip=true,width=0.8\columnwidth]{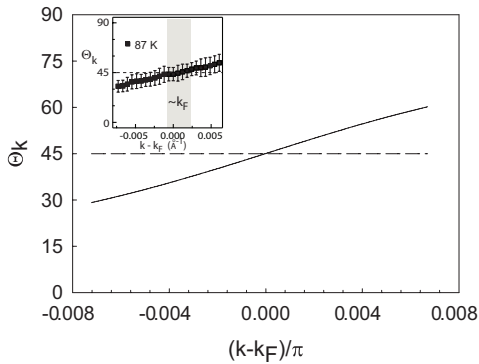}
\caption{The Bogoliubov angle along the cut position
$[0.82\pi,0.57\pi]$ to $[0.83\pi,0.58\pi]$ with $T=0.002J$ for
$t/J=2.5$ and $t'/t=0.3$ at $\delta=0.15$. Inset: the corresponding
experimental results taken from Ref. \onlinecite{balatsky}.}
\end{center}
\end{figure}

The essential physics of the momentum dependence of the
quasiparticle spectrum and the related Bogoliubov angle in cuprate
superconductors in the SC state is the same as in the case of Ref.
\onlinecite{guo}, where the doping and temperature dependence of the
low energy electron spectral function at the antinodal point are
discussed within the kinetic energy driven SC mechanism, and the
results are qualitatively consistent with the corresponding ARPES
experimental data \cite{shen,campuzano5}. Incorporating these
previous results \cite{guo}, the good agreement between the ARPES
experimental data \cite{matsui,balatsky} and the present theoretical
results within the kinetic energy driven superconductivity is
further confirmation of the conventional Bogoliubov quasiparticle
concept for cuprate superconductors.

In conclusion we have shown very clearly in this paper that the
basic d-wave BCS formalism under the kinetic energy driven SC
mechanism can correctly reproduce some low energy features found in
ARPES measurements on cuprate superconductors. Our results show that
the Bogoliubov quasiparticle intensity break near the Fermi surface
shows the existence of a gap and two branches of dispersion centered
at the Fermi surface. By calculation of the ratio of the low energy
quasiparticle spectra at positive and negative energies, we show
that the Bogoliubov angle increase monotonically across the Fermi
crossing point.

\acknowledgments

This work was supported by the National Natural Science Foundation
of China under Grant No. 10774015, and the funds from the Ministry
of Science and Technology of China under Grant Nos. 2006CB601002 and
2006CB921300.

\end{document}